\documentclass{aa}  

\usepackage{graphicx}
\usepackage{subcaption}
\usepackage{txfonts}
\usepackage{float}
\usepackage{comment}
\usepackage{newtxtext}
\usepackage[varvw]{newtxmath}
\usepackage{gensymb}
\usepackage[debug,colorlinks=true,linkcolor=blue,citecolor=blue,urlcolor=blue]{hyperref}
\usepackage{placeins}
\usepackage[linesnumbered,ruled,vlined]{algorithm2e}
\usepackage{caption}
\usepackage{setspace}

\newcommand{\doublehline}{\hline\hline\\[-1.5ex]}

\begin{document}

   \title{Direction-Dependent Faraday Synthesis}
   
\titlerunning{Direction-Dependent Faraday Synthesis}

   \subtitle{}

   \author{V. Gustafsson
          \inst{1}
          \and
          M. Br\"uggen \inst{1}
          \and
          C. Tasse \inst{2,3}
          \and
          T. En\ss lin \inst{4,5,6,7}
          \and
          S. P. O'Sullivan \inst{8}
          \and
          F. de Gasperin \inst{9}
          }

   \institute{$^1$ Hamburger Sternwarte, Universität Hamburg, Gojenbergsweg 112, 21029 Hamburg, Germany\\
   $^2$ GEPI, Observatoire de Paris, Universié PSL, CNRS, 5 Place Jules Janssen, 92190 Meudon, France\\
   $^3$ Department of Physics \& Electronics, Rhodes University, PO Box 94, Grahamstown 6140, South Africa\\
   $^4$ Max Planck Institute for Astrophysics, Karl-Schwarzschild-Str. 1, 85748 Garching, Germany\\
   $^5$ Deutsches Zentrum f\"ur Astrophysik, Postplatz 1, 02826 Görlitz, Germany\\
   $^6$ Ludwig-Maximilians-Universität M\"unchen, Geschwister-Scholl-Platz 1, 80539 Munich, Germany\\
   $^7$ Excellence Cluster ORIGINS, Boltzmannstr. 2, 85748 Garching, Germany\\
   $^8$ Departamento de Física de la Tierra y Astrofísica \& IPARCOS-UCM, Universidad Complutense de Madrid, 28040 Madrid, Spain\\
   $^9$ INAF – Istituto di Radioastronomia, via P. Gobetti 101, 40129 Bologna, Italy
   }

   \date{Received \today}

 
\abstract
{Modern radio interferometers enable high-resolution polarization imaging, providing insights into cosmic magnetism through Rotation Measure (RM) synthesis. Traditional 2+1D RM synthesis treats the 2D spatial transform and the 1D transform in frequency space separately. A fully 3D approach, transforms data directly from two spatial frequencies and one wave frequency $(u,v,\nu)$ to sky-Faraday depth space, using a 3D Fourier transform. Faraday synthesis, uses the entire dataset for improved reconstruction but also requires a 3D deconvolution algorithm to subtract artifacts from the residual image. However, applying this method to modern interferometers requires corrections for direction-dependent effects (DDEs).}
{We extend Faraday synthesis by incorporating direction-dependent corrections, allowing for accurate polarized imaging in the presence of instrumental and ionospheric effects.}
{We implement this method within \textsc{ddfacet}, introducing a direction-dependent deconvolution algorithm (DDFSCLEAN) that applies DDE corrections in a faceted framework. Additionally, we parameterize the CLEAN components, and evaluate the model onto a larger subset of frequency channels, naturally correcting for bandwidth depolarization. We test our method on both synthetic and real interferometric data.}
{Our results show that Faraday synthesis enables deeper deconvolution, reducing artifacts and increasing the dynamic range. The bandwidth depolarization correction improves the recovery of polarized flux, allowing for coarser frequency resolution without losing sensitivity at high Faraday depths. From the 3D reconstruction, we also identify a polarized source in a LOFAR surveys pointing that was not detected by previous RM surveys. Faraday synthesis is memory-intensive due to the large transforms between the visibility domain and the Faraday cube, and thus is only now becoming practical. Nevertheless, our implementation achieves comparable or faster runtimes than the 2+1D approach, making it a competitive alternative for polarization imaging.}
{}


    \maketitle
%
\section{Introduction}
\label{sec:introduction}

Modern radio interferometers, such as MeerKAT \citep{Jonas_2016}, the Australian Square Kilometre Array Pathfinder \citep[ASKAP;][]{Vanderwoude_2024}, the LOw Frequency ARray \citep[LOFAR;][]{van_Haarlem_2013} and The Expanded Very Large Array \citep[EVLA;][]{Perley_2011}, have enabled high-resolution, wide-field imaging across broad frequency ranges. Among their many applications, the study of linear polarization provides a unique window into the magnetic fields of astrophysical objects, as well as the intervening medium through which the radio waves they emit propagates \citep[e.g.]{Anderson_2024, Bockmann_2023, OSullivan_2023, De_Rubeis_2024}. Over the past two decades, Faraday rotation measure (RM) synthesis \citep{Brentjens2005} has become an important technique to study magnetic fields in general and cosmic magnetism in particular. This method decomposes the line-of-sight (LOS) Faraday rotating signal into its constituent components as a function of Faraday depth, defined as:

\begin{equation}
\label{eq:Faraday_depth}
    \phi(z) = \big(0.81\;\text{rad m$^{-2}$}\big)\int_0^z \frac{\text{d}z'}{\text{pc}}\frac{n_e(z')}{\text{cm}^{-3}}\frac{B_z(z')}{\mu \text{G}},
\end{equation}

where $n_e$ is the net number density of thermal electrons and positrons, accounting for their opposite charges, and $B_z$ is the LOS component of the magnetic field. Faraday depth quantifies the rate at which the polarization angle, $\chi$, changes as the electric field vector of electro-magnetic radiation rotates while propagating through a magnetized plasma. The amount of rotation at wavelength $\lambda$ is given by:

\begin{equation}
\label{eq:rotation}
    \chi(\lambda^2) = \chi_0 + \phi \lambda^2,
\end{equation}

where, $\chi_0$ is the intrinsic polarization angle at the source of emission. In the simplest case when there is a single Faraday rotating screen along the LOS, and no internal Faraday rotation within the emitting source, the Faraday depth is equivalent to the rotation measure. In this scenario, as described by \cite{Burn1966}, the polarization angle varies linearly with the square of the wavelength, and the RM is defined as:

\begin{equation}
\label{eq:RM}
    \text{RM}(\lambda^2) = \frac{\partial \chi (\lambda^2)}{\partial \lambda^2}.
\end{equation}

The RM synthesis algorithm is traditionally applied to each LOS independently of spatially de-convolved radio interferometric images, a technique we will refer to as 2+1D. \cite{Bell2012} introduced a 3D framework, called Faraday synthesis, that unifies radio interferometric aperture synthesis with RM synthesis. This approach transforms data directly from baseline-frequency coordinates $(u,v,\nu)$ to sky-Faraday depth space $(l,m,\phi)$, using a 3D Fourier transform. 
While a 2+1D Fourier transform is mathematically equivalent to a 3D Fourier transform, the image reconstruction by a 2+1D RM synthesis and a 3D Faraday Synthesis are generally not. The reason is that both methods apply non-linear operations to reconstruct information not measured, RM synthesis sequentially in 2D and 1D, and the Faraday Synthesis in 3D, which are not equivalent. \cite{Bell2012} therefore also introduced a novel deconvolution algorithm based on the CLEAN algorithm \citep{Hogbom_1974}, called FSCLEAN, using the full data-space sampling to construct a three-dimensional PSF, which is used to subtract artifacts from the residual image. Simultaneously, the model image is populated with point-source components in the sky-Faraday depth space, corresponding to the brightest point in the residual cube. Using a major-minor cycle scheme, as introduced by \cite{Clark_1980}, the algorithm performs a series of minor iterations in the sky domain, followed by a transformation of the predicted sky model to data space, where the next set of residual visibilities are computed. This process is repeated multiple times, progressively refining the model until convergence is achieved, ensuring alignment between the sky model and the visibility-frequency data. While this technique was able to achieve great improvements in dynamic range, spatial resolution, artifact reduction, and accurate reconstruction of low signal-to-noise sources, its adoption by the radio astronomy community has been limited. One challenge is adapting the method to modern interferometers, which require corrections for direction-dependent effects (DDEs) such as complex beam patterns and ionospheric Faraday rotation.

In this work, we extend the approach of \cite{Bell2012} to incorporate DDEs by integrating Faraday synthesis into \textsc{ddfacet} \citep{Tasse2018}, a state-of-the-art imaging tool designed for wide-field radio interferometry. \textsc{ddfacet} uses the faceting approach, introduced by \cite{van_Weeren_2016}, dividing the field of view into smaller regions and independently applying DDE corrections to each facet during the (de)gridding process. By combining Faraday synthesis with \textsc{ddfacet}, we create a unified framework that systematically reconstructs the Faraday dispersion while applying DDE corrections, enabling effective analysis of polarization data with modern interferometers. Furthermore, we demonstrate the advantages of Faraday synthesis using real observational data, a novel application to the best of our knowledge.

This paper is structured as follows: Sec.~\ref{sec_DD_Faraday_Synthesis} presents the direction-dependent Faraday synthesis technique, in terms of linear algebra. Building on the derivation in \cite{Tasse2018}, we extend the approach to incorporate RM synthesis. In Sec.~\ref{sec:deconvolution} we describe the deconvolution algorithm for Faraday synthesis. In Sec.~\ref{sec:simulation} and \ref{sec:LOFAR} we present results from applying the reconstruction method to simulated and observational data respectively, while comparing with the traditional 2+1D approach. The results are discussed in Sec.~\ref{sec:discussion} before we conclude in Sec.~\ref{sec:conclusions}.

\section{Direction-dependent Faraday synthesis}
\label{sec_DD_Faraday_Synthesis}

For a given direction $\mathbf{s}$ in the sky, the $2\times2$ visibility correlation matrix $V_{pq,(t\nu)}$ for antennas $p$ and $q$, at time $t$ and frequency $\nu$, is described by the radio interferometry measurement equation (RIME; see, e.g., \citealt{Smirnov_2011}):

\begin{equation}
\label{eq:measurement}
    V_{pq,(t\nu)} = \int_{\mathbf{s}} \bigg(\mathbf{G}_{p\mathbf{s}t\nu}\mathbf{X_s}\mathbf{G}_{q\mathbf{s}t\nu}^{H}\bigg)k^{\mathbf{s}}_{(pq),t\nu}d\mathbf{s} + n_{(pq),t\nu},
\end{equation}

\begin{equation}
\label{eq:geometry}
    \text{with } k^{\mathbf{s}}_{(pq),t\nu} = e^{-2i\pi\frac{\nu}{c}\mathbf{b}_{pq}^T(\mathbf{s}-\mathbf{s}_0)}d\mathbf{s} ,
\end{equation}
where $\mathbf{X_s}$ is the true sky electric field correlation tensor for the direction $\mathbf{s} = [l \; m \; n]^T$, $\mathbf{b}_{pq}$ is the baseline vector $[u_{pq,t} \; v_{pq,t} \; w_{pq,t}]^T$, and $n_{(pq),t\nu}$ is a random variable describing the system noise, which for the rest of this derivation we will assume is zero. $\mathbf{G}_{p\mathbf{s}t\nu}$ and $\mathbf{G}_{q\mathbf{s}t\nu}^{H}$ are the Jones matrices for antennas $p$ and $q$, respectively, where $\mathbf{G}^{H}$ is the Hermitian transpose of $\mathbf{G}$. These $2 \times 2$ matrices represent direction-dependent effects that influence the signal as it travels from the source to the antennas, such as the primary beam, ionospheric distortions, and Faraday rotation. These Jones matrices vary over time, frequency, and position in the sky.

For the rest of this paper, we will decompose Eq.~\ref{eq:measurement} into linear transformations as:

\begin{equation}
\label{eq:measurement_linear}
    \mathbf{v}_{\mathbf{b}_{\nu}} = \boldsymbol{\mathcal{S}}_{\mathbf{b}_{\nu}}\boldsymbol{\mathcal{F}}\boldsymbol{\mathcal{M}}_{\mathbf{b}_{\nu}}\mathbf{x}_{\nu} \stackrel{\text{def} \boldsymbol{\mathcal{A}}_{\mathbf{b}_{\nu}}}{=} \boldsymbol{\mathcal{A}}_{\mathbf{b}_{\nu}} \mathbf{x}_{\nu} ,
\end{equation}
where $\mathbf{v_{b_{\nu}}}$ is the visibility 4-vector, sampled by baseline $\mathbf{b}=(p,q,t)$ at frequency $\nu$, $\boldsymbol{\mathcal{M}}_{\mathbf{b}_{\nu}}$ is a $(4n_{\text{x}}) \times (4n_{\text{x}})$ block diagonal matrix representing the DDEs over the image domain, $\boldsymbol{\mathcal{F}}$ is the Fourier transform operator of size $(4n_{\text{v}}) \times (4n_{\text{x}})$, transforming from sky coordinates $(l,m)$ to visibility coordinates $(u,v)$. $\boldsymbol{\mathcal{S}}_{\mathbf{b}_{\nu}}$ is the sampling matrix of size $4 \times (4n_{\text{v}})$, which selects the 4 visibilities corresponding to $\mathbf{b}_{\nu}$. $\mathbf{x}_{\nu}$ is the true sky image vector of size $4n_{\text{x}}$.

To describe the forward mapping from the image domain to the visibility domain for all $n_{\text{v}}$ 4-visibilities associated with a specific channel $\nu$, we define the set of visibilities as $\Omega_\nu$. When $ b_\nu \in \Omega_\nu $, the index $\mathbf{b}_\nu$ spans from 1 to $n_{\text{v}}$, representing each visibility. By stacking $n_{\text{v}}$ instances of Eq.~\ref{eq:measurement_linear}, we can express the forward (image-to-visibility) mapping as:

\begin{equation}
\label{eq:measurement_freq}
\mathbf{v}_\nu = \begin{bmatrix}\vdots \\ \boldsymbol{\mathcal{A}}_{\mathbf{b}_\nu} \\ \vdots\end{bmatrix}\mathbf{x}_\nu\stackrel{\text{def}\boldsymbol{\mathcal{A}}_{\nu}}{=} \boldsymbol{\mathcal{A}}_\nu \mathbf{x}_\nu.
\end{equation}

In this formulation, $\mathbf{\mathcal{A}}_\nu$ represents the overall mapping from the image domain to the visibility domain, applying a specific direction-dependent effect (DDE) at each pixel. While pixelizing the sky introduces some approximation, $\mathbf{\mathcal{A}}_\nu$ can be considered an accurate representation of the instrument's response. As implementing $\mathbf{\mathcal{A}}_\nu$ in the forward process is computationally expensive, we introduce  $\widehat{\mathbf{\mathcal{A}}}_{\nu}$ which is the result of assuming that the DDEs in each facet can be approximated by a single $\mathbf{M}^{\mathbf{b}_{\nu}}_{\mathbf{s}_{\varphi}}$, where $\mathbf{s}_\varphi$ is direction of a subsample of the sky (facet). For the case of our derivation, we assume that the approximation is perfect, i.e. $\widehat{\mathbf{\mathcal{A}}}_{\nu} = \mathbf{\mathcal{A}}_\nu$.

So far the derivation has been exactly as \cite{Tasse2018}. Now we want to introduce RM synthesis into Eq.~\ref{eq:measurement_freq}. The first step is to add a dimension to Eq.~\ref{eq:measurement_freq}, thus not only considering a single frequency $\nu$, but instead stacking the entire set of channels. The forward mapping from image cube to visibility cube is

\begin{equation}
    \mathbf{v} = \begin{bmatrix}\vdots \\ \boldsymbol{\mathcal{A}}_{\nu} \\\vdots\end{bmatrix} \mathbf{x}\stackrel{\text{def}\boldsymbol{\mathcal{A}}}{=} \boldsymbol{\mathcal{A}} \mathbf{x}.
\end{equation}

Since RM synthesis operates on the linear Stokes parameters $Q$ and $U$, we consider only the contribution of $\mathbf{x}$ from these parameters. The conversion from the linear Stokes vector to the linear correlation products\footnote{As the linear Stokes vectors only yields the $(RL,LR)$ circular correlations, we use the linear correlations in this paper.} is achieved using the matrix $\mathbf{S}$:

\begin{equation}
    \mathbf{S} = \frac{1}{\sqrt{2}}\begin{bmatrix}
    1 & 0 \\
    0 & 1 \\
    0 & 1 \\
    -1 & 0
    \end{bmatrix}.
\end{equation}

This conversion yields the linear correlation vector $\mathbf{x}$ as:

\begin{equation}
    \mathbf{x} = \mathbf{S} \mathbf{s},
\end{equation}
where $\mathbf{s} = [Q, U]^T$ is the linear Stokes vector. We introduce the RM synthesis measurement equation

\begin{equation}
    P_{\lambda^2} = \int_{\phi}F_{\phi}e^{2i\phi\lambda^2}d\phi,
\end{equation}
which transforms the Faraday dispersion function $F_{\phi}$ to the complex polarized intensity $P_{\lambda^2}=Q_{\lambda^2} + iU_{\lambda^2}$ at wavelength $\lambda$. In terms of linear algebra, the RM synthesis forward operator $\boldsymbol{\mathcal{R}}$ is a $(2{n_{\lambda^2}}) \times (2{n_{\phi}})$ matrix, where each $(2\times2)$ block is the kernel of the RM synthesis basis $e^{2i\phi\lambda^2}$. Separating the real and imaginary terms yields the RM synthesis forward step in terms of the linear transformation $\boldsymbol{\mathcal{R}}$,

\begin{equation}
    \mathbf{s} = \boldsymbol{\mathcal{S}}_{\lambda^2}\boldsymbol{\mathcal{R}} \mathbf{s}',
\end{equation}

\begin{equation}
\text{with } \boldsymbol{\mathcal{R}}_{ij} = 
    \begin{bmatrix}
            \cos(2\phi_i\lambda^2_j) & -\sin(2\phi_i\lambda^2_j) \\
            \sin(2\phi_i\lambda^2_j) & \cos(2\phi_i\lambda^2_j)
    \end{bmatrix},
\end{equation}
where $\mathbf{s}'=[Q', U']^T$ is the Faraday dispersion with real and imaginary parts $Q'$ and $U'$ respectively and $\boldsymbol{\mathcal{S}}_{\lambda^2}$ is the sampling matrix of size $2 \times (2n_{\lambda^2})$ that selects the $n_{\lambda^2}$ sampled channels for each for each Stokes parameter. The forward step from $(l,m,\phi)$ to $(u,v,\nu)$ is then expressed as:

\begin{equation}
\label{eq:forward}
    \mathbf{v} = \boldsymbol{\mathcal{A}} \mathbf{S} \boldsymbol{\mathcal{S}}_{\lambda^2} \boldsymbol{\mathcal{R}}\mathbf{s}'\stackrel{\text{def}\boldsymbol{\mathcal{B}}}{=}\boldsymbol{\mathcal{B}}\mathbf{s}' .
\end{equation}

Forming the dirty Faraday cube involves estimating the adjoint operator $\boldsymbol{\mathcal{B}}^{H}$, which, due to factors such as calibration errors, serves as an approximation of the true adjoint operator. The dirty Faraday cube is then computed as:


\begin{equation}
\label{eq:dirty}
    \mathbf{y} = \boldsymbol{\mathcal{B}}^{H}\boldsymbol{\mathcal{W}}\mathbf{v},
\end{equation}
where $\boldsymbol{\mathcal{W}}$ is a diagonal matrix containing the set of weights for each baseline-time-frequency point.

\subsection{Per-facet 3D PSF}

As mentioned by \cite{Tasse2018}, the full polarization point spread function (PSF) is a $(4n_x) \times (4n_x)$ matrix per facet, with the diagonal terms corresponding to the $I,Q,U,V$ PSFs. Obtaining all PSFs corresponds to independently simulating four sources, one for each Stokes parameter, and computing their respective full polarization response. For example, the PSF for Stokes $I$ is computed by simulating a source with the Stokes vector $[1,0,0,0]^T$, while the PSF for Stokes Q requires $[0,1,0,0]^T$, and similarly for Stokes U and V. Instead we make the assumption that the PSF is polarization invariant, and that the Stokes $I$ PSF is a good approximation of the $|Q+iU|$ PSF. We tested this approximation by producing the PSF for each polarization for the MeerKAT L-band and found that they differ by at most only a few percent, leading us to conclude that this approximation is sufficiently accurate for practical purposes. However, further investigation is required to determine whether this approximation remains valid for LOFAR, where the linear feeds are not necessarily orthogonal to the pointing center, and their orientations vary between stations, potentially introducing larger systematic differences. To minimize the potential impact of using an incorrect PSF, we implement early stopping in the deconvolution minor cycles, and trust that any errors are corrected for in the major cycle.

The 3D PSF is defined as the response of the system to a polarized point source of unit amplitude located at the location $(l,m,\phi)$. However, since we are approximating the polarization PSF with the Stokes $I$ PSF, the backwards step is not simply the Hermitian transpose of the forward step. Instead, the 3D PSF $\mathbf{y_1}$ is computed as

\begin{equation}
\label{eq:PSF}
    \mathbf{y_1} = \boldsymbol{\mathcal{B}}^H\boldsymbol{W}\boldsymbol{\mathcal{A}}\mathbf{S}\mathbf{0_1},
\end{equation}
where $\mathbf{0_1}$ is a vector containing zeros everywhere, except the central pixel, which is set to $[I,Q,U,V]=[1,0,0,0]$ for each frequency channel. This operation is performed for each facet, applying the facet dependent $\boldsymbol{\mathcal{M}}$, and its Hermitian transpose during the forward and backward step, respectively.

\section{Deconvolution}
\label{sec:deconvolution}

In this section, we describe the deconvolution method used for Faraday synthesis. We introduce modifications to the approach developed by \cite{Bell2012} to better integrate it into the \textsc{ddfacet} framework, resulting in Direction-Dependent Faraday Synthesis CLEAN (DDFSCLEAN). The algorithm is outlined in Alg.~\ref{alg:DDFSCLEAN}, and we provide a more detailed explanation below.

The algorithm is initialized by averaging the visibilities from the measurement set across the full set of frequency channels $\nu_{\text{MS}}$ down to the gridding frequency channels $\nu_{\text{Grid}}$. The dirty Faraday cube is then constructed according to Eq.~\ref{eq:dirty}, along with the PSF, as described in Eq.~\ref{eq:PSF}. The noise level $\sigma_{QU}$ is calculated from

\begin{equation}
    \sigma_{QU}^2 = \frac{\sigma_{Q}^2 + \sigma_{U}^2}{2},
\end{equation}

\begin{equation}
   \text{with } \sigma_Q^2 = \frac{1}{N} \sum_i^N \sigma_{Q,i}^2, \quad
   \sigma_U^2 = \frac{1}{N} \sum_i^N \sigma_{U,i}^2,
\end{equation}
where $N$ is the number of gridding channels. This noise level is used as a final threshold for the cleaning process. However, due to the positive bias of the Rice distribution, the false detection rate is higher compared to Gaussian statistics. As a result, a higher threshold in terms of $\sigma_{QU}$ must be used to mitigate spurious detections \citep{George_2012}. Once initialized, the algorithm identifies the brightest point in the polarized intensity cube and performs a least-squares fit between the peak Faraday dispersion and a 1D Gaussian profile, $\mathcal{G}(\phi)$. The best-fit parameters, with the flux scaled by the loop gain, $\alpha$, and corrected for Rician bias following \cite{George_2012}, along with the phase interpolated to the corrected Faraday depth, are then added to the model. The operator $\mathbf{\Theta}$ maps the component parameters onto the Faraday cube and applies a convolution with the facet PSF. The resulting component is then scaled by the loop gain and subtracted from the residual image. This iterative procedure continues until a user-defined stopping criterion is met, either when the peak flux falls below a specified threshold, $t$, or when the maximum number of minor iterations is reached. 
At this point the model parameters are evaluated onto the degridding frequencies $\nu_\text{Degrid}$  
and forward-mapped to data space according to Eq.~\ref{eq:forward}, while applying direction-dependent Jones matrices, as well as correcting for bandwidth depolarization, see Appendix~\ref{sec:Bandwidth}. 
In data space, the residual visibilities are computed as the difference between the observed data and the modeled visibilities. These residuals are then mapped back to sky-Faraday depth space (Eq.~\ref{eq:dirty}), forming a new residual cube that is used to drive the next cycles of deconvolution. Once the residual peak flux is below the final threshold, the deconvolution process is terminated. The model cube is constructed by applying the operator $\boldsymbol{\Pi}$ which maps the model parameters onto an empty Faraday cube. The restored cube is then obtained by convolving the model cube with a 3D Gaussian to match the desired resolution, represented by the convolution operator $\boldsymbol{\mathcal{C}}$, and adding it to the final residual Faraday cube.

\begin{algorithm}
\setstretch{1.2}
\caption{\textbf{DDFSCLEAN}}
\label{alg:DDFSCLEAN}
\KwData{$\mathbf{v}$, $\mathbf{y}$, $t$, $\alpha$}
\SetKwInOut{Initialization}{Initialization}
\Initialization{$\widehat{\boldsymbol{\pi}} = \mathbf{0};$}
\CommentSty{/* Start $n_\mathrm{Cycle}$ major cycles */}\;
\For{$i_\mathrm{Cycle} \in \mathrm{range}(n_\mathrm{Cycle})$}{
    \CommentSty{/* Start minor cycles */}\;
    \While{$\mathrm{max}\{|\mathbf{y}|\} > t$}{
        Find location of brightest pixel\;
        $i = \text{argmax}\{|\mathbf{y}|\}$\;
        Find locally best sky model centered on pixel $i$\;
        $(A_i,\phi_i,\chi_i) = \text{argmin}_{\boldsymbol{\pi}_i} \left||\mathbf{y}_i| - \mathcal{G}(\boldsymbol{\pi}_i) \right|^2$\;
        Update sky model:\\ 
        $\widehat{\boldsymbol{\pi}} \gets \widehat{\boldsymbol{\pi}} + 
        (\alpha A_i,\phi_i,\chi_i)$\;
        Update residual cube:\\
        $\mathbf{y} \gets \mathbf{y} - \mathbf{\Theta} (\alpha A_i,\phi_i,\chi_i)$\;
    }
    Update residual cube:\\ $\mathbf{y} = \boldsymbol{\mathcal{B}}^{H}\boldsymbol{\mathcal{W}}(\mathbf{v} - \boldsymbol{\mathcal{B}}\widehat{\boldsymbol{\pi}})$\;
}
Compute model cube: $\widehat{\mathbf{x}} = \boldsymbol{\Pi}\widehat{\boldsymbol{\pi}}$\;
Compute restored cube: $\widehat{\mathbf{x}} \gets \mathbf{y} + \boldsymbol{\mathcal{C}}\widehat{\mathbf{x}}$
\end{algorithm}

\section{Simulation}
\label{sec:simulation}

In order to test direction-dependent Faraday synthesis under observationally representative conditions, we use the MeerKAT L-band measurement set from \cite{de_Gasperin_2022}, replacing the visibilities with our simulation while keeping the original flagging. A summary of the observation configuration can be seen in Table~\ref{tab:config}. 

\begin{table}
    \centering
    \caption{Observation configuration of synthetic MeerKAT data.}
    \begin{tabular}{@{}p{0.5\columnwidth} p{0.3\columnwidth}@{}}
    \doublehline
       Field of view & $1\degree \times 1\degree$ \\
       Frequency range & 856 - 1712 MHz\\
       -- Bandwidth & 856 MHz \\
       Frequency resolution & \\
        -- measurement set $\delta \nu_{\text{MS}}$ & 899.2 kHz \\
        -- degridding $\delta \nu_{\text{Degrid}}$ & 4.280 MHz \\
        -- gridding $\delta \nu_{\text{Grid}}$ & 17.12 MHz \\
        Maximum Faraday depth & $\pm 750$ rad m$^{-2}$ \\
        Faraday depth sampling & 4.308 rad m$^{-2}$ \\
       \hline
    \end{tabular}
    \label{tab:config}
\end{table}

The visibilities are produced using the forward mapping of \textsc{ddfacet}, incorporating time-baseline-frequency direction-dependent beam effects\footnote{The beam effects are simulated using \hyperlink{blue}{https://github.com/ratt-ru/eidos?tab=readme-ov-file}}. It consists of 100 point sources randomly distributed over a $1\degree \times 1\degree$ field of view (FOV). The Stokes $I$ fluxes are drawn from an inverse gamma distribution, which has a wide tail that allows for extremely bright sources. The emission is assumed to be spectrally flat across the observed frequency range. The fractional polarizations are limited to a maximum of 70\%, resulting in a range from 0.26 mJy to 2.69 Jy. Rotation measures are restricted to $|700|$ rad m$^{-2}$, and the intrinsic polarization angles are uniformly drawn from the right-half plane. Gaussian white noise is added to the visibilities, scaled by the square root of the number of visibilities and channels, which, depending on the weighting scheme, results in an RMS of 1 mJy beam$^{-1}$ before primary beam correction.\\

For the 2+1D deconvolution, we follow the standard approach for cleaning polarization data, while accounting for DDEs. In order to have roughly the same resolution across the bandwidth we first apply a sigmoid taper in the $uv$-plane, resulting in a resolution of $9.6\arcsec \times 9.1\arcsec$. We use Briggs weighting throughout this work with a robust factor of 0. While the resolution could be improved for both reconstruction methods by adjusting the weighting scheme, optimizing resolution is not the focus of this work, and we do not investigate it further.

The imaging begins with a 2D deconvolution driven by the peak polarized intensity map, $|P|$, defined as

\begin{equation}
    |P| = \frac{1}{N}\sum_i^N \sqrt{Q_i^2 + U_i^2},
\end{equation}
where $Q_i$ and $U_i$ are the linear polarization components at channel $i$ and $N$ is the number of gridding channels. In each minor iteration, the flux corresponding to the brightest pixel in the peak polarized map, scaled by the loop gain is added to the model cube, while the scaled and shifted spectral PSF is subtracted from the residual image. When the peak flux reaches the minor loop threshold, each facet of the model cube is Fourier transformed to data space where it is compared with the data, computing the next residual. Once the final threshold is reached, each channel of the model cube is convolved with a common restoring beam, corresponding to the lowest resolution of the bandwidth. To assess the impact of time-dependent beam effects, we also performed deconvolution without correcting for beam variations, applying only the mean beam afterward. However, this yielded no significant difference from the previous 2+1D reconstruction, so we do not present these results here.

For RM synthesis, the absolute RM search range is set based on the mean channel width in $\lambda^2$, resulting in a minimum sensitivity of $\sim 70 \%$ at the edges of the sampled Faraday depth. The Faraday depth is sampled at 10 points per FWHM of the RM synthesis dirty beam. RM synthesis and RMCLEAN are run using the \textsc{rm-tools} package, with a cleaning threshold of $5\sigma_{QU}$. Uniform weighting are used for RM synthesis, both for the 2+1D and 3D reconstructions.

The 3D reconstruction is created following the procedure described in Sec.~\ref{sec:deconvolution}. We use \textsc{rm-tools} to perform RM synthesis, as it uses the fast Fourier transform, which significantly accelerates the process. From a Gaussian fit to the real part of the main lobe of the 3D PSF, we obtain a restoring beam FWHM of $8.9\arcsec \times 8.6\arcsec \times 39.0$ rad m$^{-2}$.

Both deconvolution algorithms were run for a total of ten major iterations, with a maximum of 10000 minor iterations. We adopt early stopping in the minor cycles, stopping at half the initial peak flux. After deconvolution, we fit a 3D Gaussian to peaks at the positions of the true signals, using the peak flux and corresponding RM as initial estimates. The intrinsic polarization angle is recovered by linear interpolation between the adjacent sampling in $\phi$ and the flux is corrected for the polarization bias following \cite{George_2012}.\\

\begin{figure*}
    \centering
    \includegraphics[width=\textwidth]{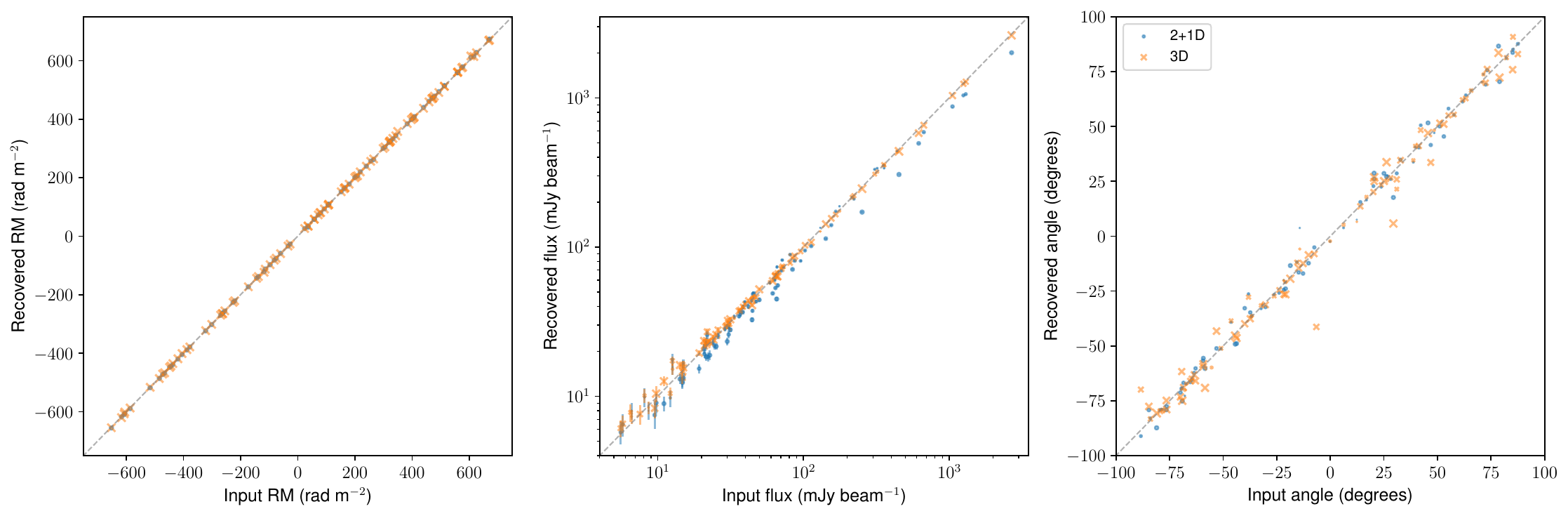}
    \caption{Reconstructed source parameters compared to the input of the synthetic MeerKAT dataset. Blue dots denote the 2+1D reconstruction while orange crosses denote the 3D reconstruction. Left: True vs. recovered RM. Middle: True vs. recovered flux. Right: True vs. recovered intrinsic polarization angle. Marker sizes are proportional to the absolute RM.}
    \label{fig:Input_vs_Output}
\end{figure*}

In Fig.~\ref{fig:Input_vs_Output}, we compare the 2+1D and 3D deconvolution algorithms in terms of their ability to recover the input parameters. Detections with a signal-to-noise ratio below four are excluded from the plot to minimize false positives. Both methods successfully recover the input RM, with no significant difference between them. The 3D reconstruction shows an accurate recovery of the flux across the dynamic range, with no significant correlation with the input RM. In contrast, the 2+1D deconvolution systematically underestimates the flux of high-RM components, due to bandwidth depolarization, as indicated by the marker sizes in the figure. As for the polarization angle both methods show errors centered at zero, with mean absolute errors of $3.2\degree$ and $3.4\degree$ for the 2+1D and 3D algorithms respectively.

\begin{figure}
    \centering
    \includegraphics[width=\columnwidth]{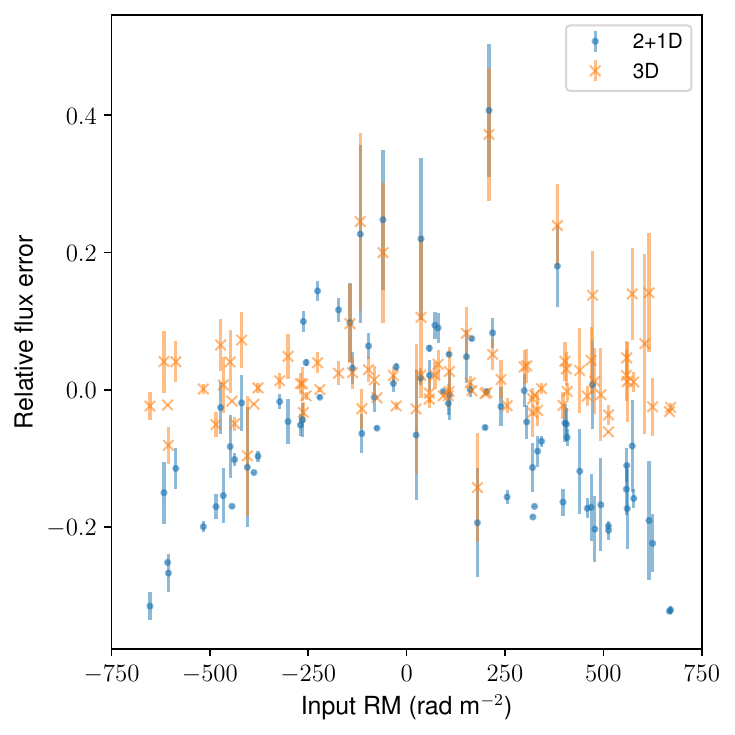}
    \caption{A comparison between the 2+1D (blue) and 3D (orange) methods on the effects of bandwidth depolarization. Only signals with a signal to noise above $8\sigma_{QU}$ are included in the figure.}
    \label{fig:Depolarization}
\end{figure}

Fig.~\ref{fig:Depolarization} shows a better view of the bandwidth depolarization effect of channel averaging. We see that by not correcting for bandwidth depolarization, the recovered flux exhibits a systematic trend with Faraday depth. The 2+1D method underestimates flux at high Faraday depths, while the bandwidth corrected 3D reconstruction show errors mostly centered at zero. The errorbars generally appear to increase as a function of $|\phi|$, due to the lower signal to noise level at the lower sensitivities. In Appendix~\ref{sec:Bandwidth} we attempt to correct the 2+1D reconstruction for bandwidth depolarization, using a tool from the \textsc{rm-tools} package.

\begin{figure}
    \centering
    \includegraphics[width=\columnwidth]{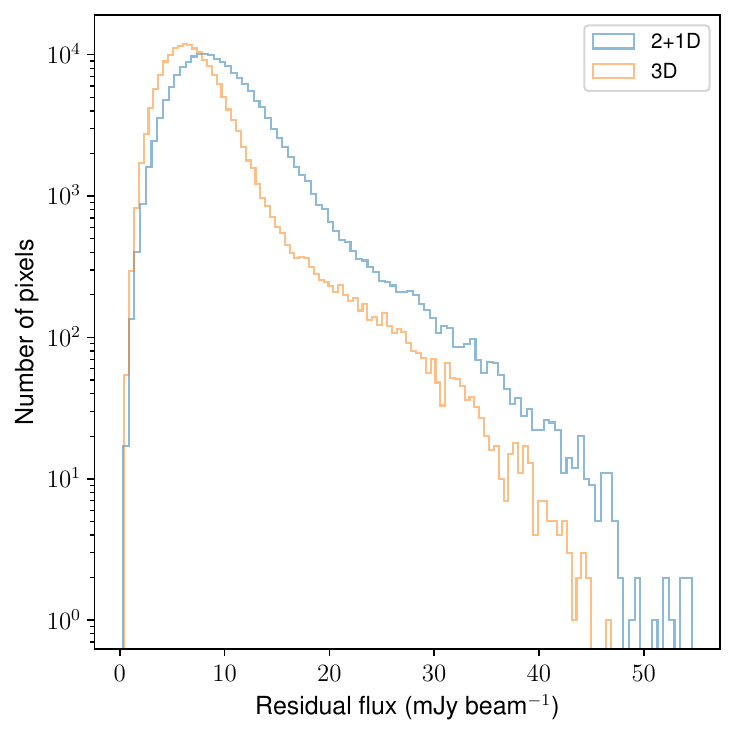}
    \caption{The flux distribution in the residual spectral cube for the 2+1D (blue) and 3D (orange) deconvolution methods.}
    \label{fig:Residual}
\end{figure}

In order to evaluate the performance of the two deconvolution methods, independently of bandwidth depolarization correction, we examine the distribution of flux in the residual spectral cube, as shown in Fig.~\ref{fig:Residual}. The 2+1D reconstruction shows a clear shift in the peak position of the distribution, indicating a higher residual root mean square. Additionally, the presence of higher flux values is a result of the shallower deconvolution, leaving more signal flux in the residuals.

In Table~\ref{tab:runtime} we show the runtimes of the 2+1D and 3D imaging techniques. Both algorithms are run on an AMD EPYC 7H12 CPU, running at a clock frequency of 2.6 GHz, using up to 128 threads for parallelization, and a maximum of 500 GB of memory. The main bottleneck of the 3D deconvolution is the degridding step onto the larger set of frequency channels. The minor iterations contribute only a small fraction of the total computational cost. As a result, although the 2+1D deconvolution requires more iterations, this does not significantly impact its total runtime. If we add up the runtimes of the 2+1D method, the total is longer than for the 3D method. The 3D deconvolution can be further accelerated by reducing the number of degridding channels, at the cost of less accurate bandwidth depolarization correction.\\

\begin{table}
    \centering
    \caption{Runtimes on synthetic MeerKAT data.}
    \label{tab:runtime}
    \renewcommand{\arraystretch}{1.2} 
    \begin{tabular}{@{}p{0.6\columnwidth} p{0.2\columnwidth}@{}}
    \doublehline
    \textbf{Method} & \textbf{Runtime} \\
    \hline
    3D & 3h36m \\
    \hline
    2+1D & 3h49m \\
    -- Aperture synthesis + CLEAN & 3h0m \\
    -- RM synthesis & 38m \\
    -- RMCLEAN & 11m \\
    \hline
    \end{tabular}
\end{table}

\section{LOFAR observation}
\label{sec:LOFAR}

In this section, we apply our implementation of the 3D imaging technique to a LOFAR HBA pointing from LoTSS DR2 \citep{Shimwell_2022}. From the catalog of polarized sources produced by \cite{OSullivan_2023} (hereafter referred to as \textbf{OS23}), we select the region with the highest density of polarized sources per square degree and use the corresponding pointing for our analysis. The imaging configuration of the dataset is shown in Table~\ref{tab:LOFAR}. We use the full frequency resolution, both, in image and in data space, which results in a maximum observable Faraday depth of $\pm 170$ rad m$^{-2}$ at full sensitivity and $\pm 450$ rad m$^{-2}$ at 50\% sensitivity. We use the same sampling as \textbf{OS23}, with a maximum Faraday depth of $\pm120$ rad m$^{-2}$ and a sampling of $0.05$ rad m$^{-2}$, resulting in 4801 samples in $\phi$. As our goal is to produce a polarized intensity map and an RM map, we use the \textsc{rmsynth3d} function from \textsc{rm-tools} for 2+1D RM synthesis. Due to the varying noise level across the field, as a consequence of the primary beam correction, a 3D noise map would be required to use the inverse-variance weighting. Since this functionality is not yet implemented in either \textsc{rm-tools} or our software, we instead apply uniform weighting for RM synthesis. An attempt was made to taper the visibilities for the 2+1D method, but this introduced severe artifacts that were not removed during deconvolution. Instead, we apply a common restoring beam, set by the lowest frequency, with a resolution of $7.42\arcsec$. As a result, the residuals will have varying resolutions across the bandwidth, potentially leading to inaccurate RM estimates. For the 3D method, we fit a Gaussian to the central PSF facet, resulting in a resolution of $6.31\arcsec$. As we are using the same frequency resolution in image space as in data space, we do not apply any correction for bandwidth depolarization. We divide the FOV into $7\times7$ facets, and apply the direction-dependent calibration solutions, as well as the primary beam correction, during imaging. Due to the higher sampling in $\phi$, we run the code on a high-memory computing cluster, on a single node with 3 TB of available RAM.

\begin{table}
    \centering
    \caption{Observation configuration of LOFAR data.}
    \begin{tabular}{@{}p{0.4\columnwidth} p{0.4\columnwidth}@{}}
    \doublehline
       Pointing & P142+42 \\
       Phase center & 9h27m33s, +41\degree48\arcmin24\arcsec \\
       Field of view & $1\degree \times 1\degree$ \\
       Frequency range & 120 - 168 MHz \\
       -- Bandwidth & 48 MHz \\
       Frequency resolution & 97.6 kHz \\
       Maximum Faraday depth & $\pm 120$ rad m$^{-2}$ \\
       Faraday depth sampling & 0.05 rad m$^{-2}$ \\
       \hline
    \end{tabular}
    \label{tab:LOFAR}
\end{table}

\begin{figure*}
    \centering
    \includegraphics[width=\textwidth]{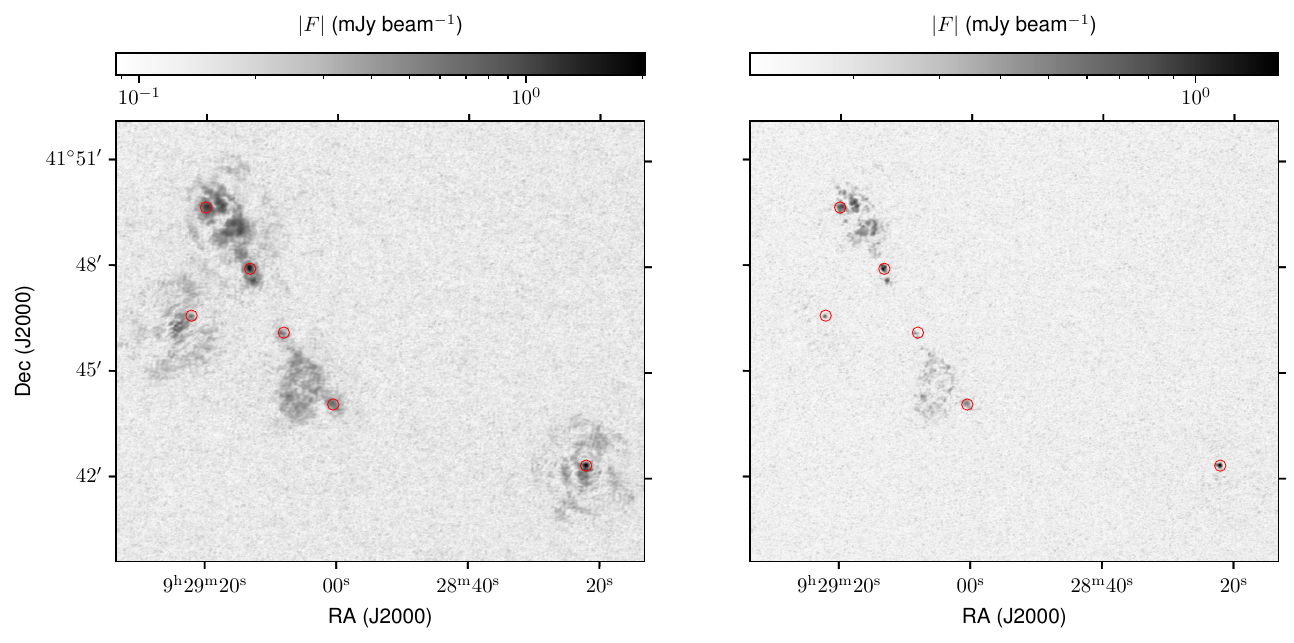}
    \caption{The 2+1D (left) and 3D (right) reconstructions for six polarized sources from \textbf{OS23}, marked by red circles. Two separate colorbars are used since the beam size differ between the two reconstructions. The 2+1D reconstruction contains significant artifacts around the northern lobe and isolated unresolved sources.}
    \label{fig:LOFAR}
\end{figure*}

In Fig.~\ref{fig:LOFAR}, we show a zoom-in on the polarized sources detected by \textbf{OS23}, comparing the 2+1D and 3D deconvolution methods. Although the sources were recorded as six separate entries in the catalog, we identify four as originating from the same double-lobed galaxy. The most notable difference between the reconstructions is the presence of artifacts around bright sources in the 2+1D case, which are barely noticeable in the 3D reconstruction. The presence of bright Stokes $I$ sources within the FOV introduces leakage into the linear polarization components, delaying the deconvolution of polarized sources until later stages. The 2+1D method struggles to recover faint emission near the noise level, which leads to incomplete deconvolution of the sources shown here. While the flux appears brighter in the 2+1D image, this can be attributed to the larger beam size together with undeconvolved extended emission, which also leads to smoother emission and blends small-scale structures. In contrast, the 3D reconstruction better preserves compact features and reduces contamination from sidelobes, resulting in a more accurate representation of the underlying source structure. We do not observe a significant difference in noise levels between the reconstructions, as both images have a $\sigma_{QU}$ of $45$ $\mu$Jy beam$^{-1}$ in regions where no sources are present.

\begin{figure*}
    \centering
    \includegraphics[width=\textwidth]{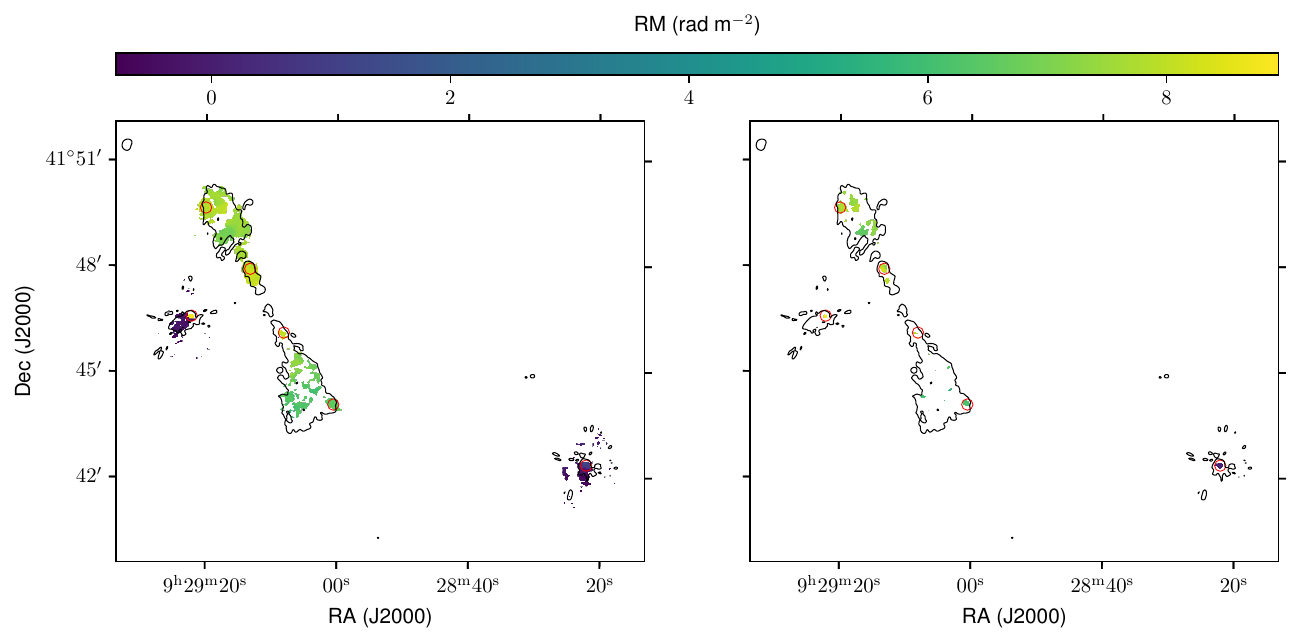}
    \caption{RM maps from the 2+1D (left) and 3D (right) reconstructions, centered on the six polarized sources from \textbf{OS23}, indicated as red circles. Only pixels with a flux higher than $8\sigma_{QU}$ are shown. The Stokes $I$ $10\sigma_I$ noise level is shown as a black contour.}
    \label{fig:LOFAR_RM}
\end{figure*}

In Fig.~\ref{fig:LOFAR_RM}, we show the RM maps from the 2+1D and 3D reconstructions. The RM is obtained from a Gaussian fit to the peaks above $5.5\sigma_{QU}$ in the Faraday cube. Only pixels with flux above $8\sigma_{QU}$, after correcting for polarization bias, are included in the final RM map. While the 2+1D recovers more flux above the $8\sigma_{QU}$ level, some of the artifacts around the compact sources are also above this level, resulting in false detections. In contrast, with the 3D approach, we can lower the detection threshold to $6.5\sigma_{QU}$ without introducing spurious detections. However, since this would not be a fair comparison, we do not include it here. For the 3D reconstruction, we see that all sources found by \textbf{OS23} are above the $8\sigma_{QU}$ level, although with a slight shift in position for some of the sources, likely due to the different resolution.

In Table~\ref{tab:comparison} we compare the recovered RMs from the two reconstructions with those reported by \textbf{OS23}. The RM is computed from a Gaussian fit to the highest peak in $\phi$, located at the positions recorded by \textbf{OS23}. The errors are calculated in the standard manner as the FWHM of the PSF in $\phi$ divided by twice the signal to noise ratio. We do not compare the Galactic RM-corrected values, as the correction procedure would be identical for both our work and \textbf{OS23}, adding no additional information. 

We find that only one RM value from the 3D reconstruction is consistent with those reported in \textbf{OS23}. The discrepancy between the values is not unexpected given the differences in resolution—both spatially and in $\phi$—as \textbf{OS23} applies inverse-variance weighting in frequency, while we have weighted each channel uniformly. The 2+1D reconstruction generally agrees more with \textbf{OS23}, with the exception of one source $(9h28m22s, +41\degree42\arcmin22\arcsec)$, which we see in Fig.~\ref{fig:LOFAR_RM} is significantly contaminated by artifacts. Furthermore, the RM is close to the Stokes $I$ leakage at $\phi=0$, potentially causing the shift to a lower recovered RM. We also compare the flux densities with those from \textbf{OS23}. The fluxes from \textbf{OS23} are systematically higher than those in our study, likely because the larger beam size integrates over a broader region, capturing more extended emission. Furthermore, the positions are slightly shifted from those identified in the 2+1D and 3D approaches due to differences in resolution. While shifting to the positions detected by our method increases the peak flux, the resulting RM values do not align more closely with those reported by \textbf{OS23}.

\begin{table*}
    \centering
    \caption{A comparison of the RM and flux values obtained from the two reconstruction methods for the six polarized sources identified in \textbf{OS23}.}
    \label{tab:comparison}
    \renewcommand{\arraystretch}{1.2} 
    \resizebox{\textwidth}{!}{ 
    \begin{tabular}{l c c c c c c}
    \hline\hline
    \textbf{Coordinate} (RA, Dec) & \multicolumn{3}{c}{\textbf{RM} (rad m$^{-2}$)} & \multicolumn{3}{c}{$\boldsymbol{\mathit{|F|}}$ (mJy beam$^{-1}$)} \\
    & \textbf{OS23} & \textbf{2+1D} & \textbf{3D} & \textbf{OS23} & \textbf{2+1D} & \textbf{3D} \\
    \hline
    (9h28m22s, +41\degree42\arcmin22\arcsec) & $1.41\pm0.01$ & $0.88\pm0.02$ & $1.37\pm0.02$ & $2.60\pm0.05$ & $1.37\pm0.05$ & $0.97\pm0.05$ \\
    (9h29m01s, +41\degree44\arcmin05\arcsec) & $6.54\pm0.02$ & $6.53\pm0.05$ & $6.53\pm0.05$ & $1.03\pm0.04$ & $0.59\pm0.05$ & $0.43\pm0.04$ \\
    (9h29m08s, +41\degree46\arcmin07\arcsec) & $8.10\pm0.03$ & $8.10\pm0.06$ & $8.21\pm0.06$ & $0.91\pm0.05$ & $0.51\pm0.05$ & $0.35\pm0.04$ \\
    (9h29m13s, +41\degree47\arcmin55\arcsec) & $8.06\pm0.01$ & $8.09\pm0.02$ & $8.38\pm0.02$ & $2.09\pm0.05$ & $1.70\pm0.04$ & $1.12\pm0.05$ \\
    (9h29m20s, +41\degree49\arcmin39\arcsec) & $7.84\pm0.01$ & $7.89\pm0.03$ & $7.91\pm0.03$ & $2.32\pm0.05$ & $1.10\pm0.04$ & $0.70\pm0.05$ \\
    (9h29m22s, +41\degree46\arcmin35\arcsec) & $8.78\pm0.03$ & $8.93\pm0.05$ & $8.91\pm0.04$ & $0.81\pm0.05$ & $0.63\pm0.05$ & $0.57\pm0.04$ \\
    \hline
    \end{tabular}
    }
\end{table*}

\begin{figure}
    \centering
    \includegraphics[width=\columnwidth]{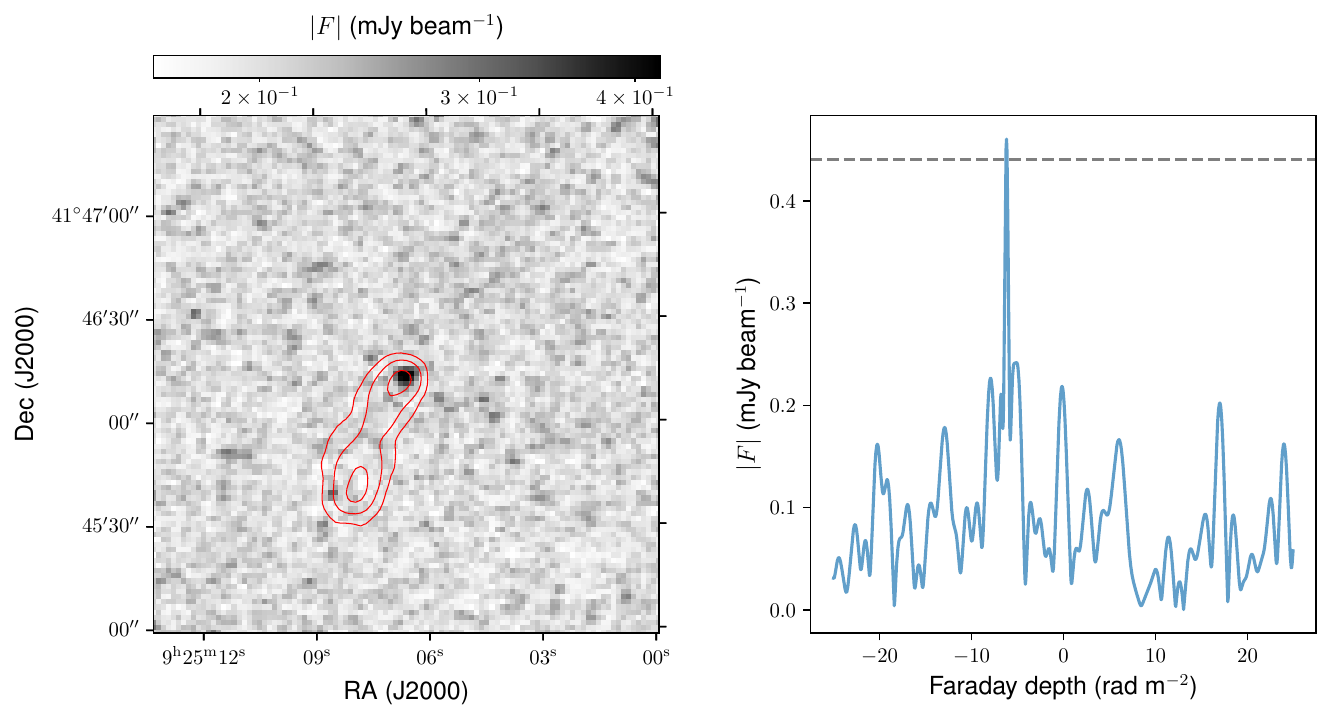}
    \caption{A polarized source previously not detected by \textbf{OS23}, found by our 3D reconstruction method. Left: Polarized intensity map centered on the source. The Stokes $I$ $(10, 20, 50)\times\sigma_{I}$ noise levels are shown as red contours. Right: Magnitude of the LOS profile at the brightest pixel. The local $8\sigma_{QU}$ level is shown as a dashed black line.}
    \label{fig:New_source}
\end{figure}

In Fig.~\ref{fig:New_source} we show a polarized source, not detected by \textbf{OS23}. The source is located $0.75\degree$ west of the sources in Fig.~\ref{fig:LOFAR}, and from the Stokes I image appears to be a double-lobed galaxy. However, only one of the lobes is detected in polarized emission outside the Stokes $I$ leakage range. From a Gaussian fit to the main peak, we obtain a RM of $-6.20\pm0.05$, in contrast to positive RMs of the other sources in the field. Whether this is due to foreground Faraday rotation, intrinsic properties of the source, or its surroundings would require further analysis. Since \textbf{OS23} determines the noise level from the wings of the dirty Faraday dispersion ($|\phi|>100$ rad m$^{-2}$) it is likely an over-estimate of the noise, particularly for bright or high-RM sources. This could explain why their search algorithm does not detect this source.

\section{Discussion}
\label{sec:discussion}

Since the 2+1D deconvolution is driven by the mean flux map from the entire bandwidth, instead of convolving a series of Stokes $Q$ and $U$ images, we do not observe the same source flux underestimation as in \cite{Bell2012}, where the reconstruction of low-flux sources was limited by the noise in individual frequency channels. Instead, the observed bias in Fig.~\ref{fig:Input_vs_Output} is not dependent on source flux but rather on the input RM, as a result of bandwidth depolarization. Correcting for bandwidth depolarization could, in principle, be applied to the 2+1D reconstruction. However, such a correction would require additional information that is typically not well known after deconvolution, making it more of an approximation than a reliable recovery of the true depolarization. As these corrections are not widely implemented to our knowledge, our main comparison in Sec.~\ref{sec:LOFAR} do not apply such a correction. However, in Appendix~\ref{sec:Bandwidth} we evaluate the accuracy of the adjoint transform correction, presented in \cite{Fine_2023}.

Our 3D reconstruction inherently accounts for the effects of channel averaging during imaging, incorporating flagged channel information, and the mapping between frequency resolutions, to determine their relative bandwidth depolarization. This approach becomes particularly advantageous in the later stages of deconvolution, where false positives amplified by the correction are naturally corrected for in the major cycle. We realize that this correction can only be applied if the frequency resolution of the data is higher than that of the image products. In the case of a LOFAR observation, where essentially the full frequency resolution is required to have a high sensitivity at high Faraday depths, alternative approaches must be used, such as those presented in \cite{Fine_2023}.

While the improvements achieved by our reconstruction technique for simulated unresolved sources are primarily due to bandwidth depolarization correction rather than Faraday synthesis alone, its true advantages become evident for spatially resolved structures, as shown in Fig.~\ref{fig:LOFAR}. The large number of degrees of freedom $(2N_{\nu_{\text{Grid}}})$ in the 2+1D reconstruction leads to an overfit of the true source flux, allowing noise and artifacts from the PSF of nearby pixels to be incorporated into the model. In contrast, the 3D reconstruction is parameterized only by the Gaussian fit parameters and the intrinsic polarization angle of the source, enabling a more stable, faster, and deeper deconvolution. This deeper deconvolution improves the accuracy of polarized emission reconstruction and increases the dynamic range, potentially enhancing RM grid densities by detecting more polarized sources.

Following \cite{Bell2012}, we have assumed that Faraday synthesis eliminates the need for resolution matching, as spectral variations in resolution across the bandwidth are inherently accounted for in the 3D PSF. However, this assumption is only valid if the source structure can be well approximated by a single point, as assumed in the CLEAN algorithm. If this condition is not met, resolution matching is still necessary unless the magnetic field is ordered and the RM is approximately constant within the beam area. Otherwise, there will be frequency-dependent beam depolarization, due to the varying resolution, which is not modeled by the 3D PSF. As our implementation of Faraday synthesis uses the CLEAN algorithm for deconvolution, we have assumed that the first condition is met and thus do not perform resolution matching. However, for sources with extended emission or complex Faraday dispersions, this assumption may lead to inaccuracies due to unmodeled frequency-dependent beam depolarization. Future studies should investigate how such effects influence the reconstruction of Faraday dispersions and whether incorporating resolution matching into the deconvolution process improves accuracy.

Furthermore, the 3D PSF constructed in this work assumes a flat spectrum across the bandwidth. In contrast, a common approach in RM synthesis is to account for the spectral dependence by either applying a spectral index correction or performing RM synthesis on the fractional polarizations,

\begin{equation}
    q_{\lambda^2} = \frac{Q_{\lambda^2}}{I_{\lambda^2}}, \quad u_{\lambda^2} = \frac{U_{\lambda^2}}{I_{\lambda^2}},
\end{equation}

where $I_{\lambda^2}$ is the modeled Stokes $I$ flux, excluding noise. However, since \textsc{ddfacet} jointly deconvolves all Stokes parameters, this correction cannot be applied within the current framework. Future work will explore incorporating a user-defined spectral index across the full field, which may improve RM recovery for sources with steep spectra.

The main challenge of Faraday synthesis is the high memory demand associated with large transformations. In particular, the Fourier transform from $\lambda^2$ to $\phi$ is the most computationally expensive step, making it impractical for large fields of view and high spectral resolution data. As a result, the LOFAR example presented in this work was limited to a field of view of $1\degree \times 1\degree$, as larger images would have exceeded memory constraints, even with 3 TB of available RAM. The memory impact can be reduced by, for example, computing only a small patch of the PSF in $\phi$, similar to what is typically done in CLEAN. While \textsc{ddfacet} generally processes each facet in parallel, memory-intensive operations can be run sequentially, though at a significant cost in computation time. The majority of the computational time for LOFAR polarization deconvolution is spent on deconvolving polarization leakage, characterized by a peak around $\phi=0$. Future studies will explore the possibility of excluding this leakage range, as it does not contribute to scientific analysis, and would otherwise be blanked as in \textbf{OS23}. This would significantly reduce the computation time for LOFAR data, as only a fraction of the current amount of minor iterations would be necessary for a given deconvolution threshold.

One potential application for direction-dependent Faraday synthesis is the The International LOFAR Two-metre Sky Survey\footnote{\hyperlink{blue}{https://lofar-surveys.org/ilotss.html}} (ILoTSS), which will deliver high-resolution ($0.3\arcsec$) data products of radio sources across the northern sky. To accurately analyze these data, polarization deconvolution will be essential for a correct recovery of the RMs and magnetic field structure in these well-resolved sources with complex magnetoionic environments.

While the MeerKAT L-band offers a spatial resolution comparable to that of LoTSS, its channel width in $\lambda^2$ is significantly narrower. This results in a much higher maximum observable Faraday depth, given by

\begin{equation} 
\label{eq:RM_sensitivity} 
    |\phi|_{\text{max}} \propto \frac{\sqrt{3}}{\delta\lambda^2}, 
\end{equation}

for a given number of frequency channels. Consequently, only a few tens of channels are needed to achieve sensitivity to Faraday depths relevant to most cosmic magnetism studies. With access to high-memory computing clusters, Faraday synthesis could also be applied to ASKAP cosmic magnetism studies \citep[POSSUM;][]{Vanderwoude_2024}, potentially increasing the RM grid density. Combined with the bandwidth depolarization correction implemented in this work, Faraday synthesis becomes a viable approach for high-resolution polarization imaging.

\section{Conclusions}
\label{sec:conclusions}

We have incorporated Faraday synthesis into \textsc{ddfacet}, a modern radio imaging tool designed for wide-field, wide-band interferometric imaging. We have also introduced a novel implementation of a direction-dependent Faraday synthesis deconvolution algorithm, and tested the algorithm on, both, synthetic data, modelled on MeerKAT observations, as well as observational data from LOFAR. The 3D deconvolution method shows a significant reduction in artifacts, while increasing both the dynamic range and resolution. The computational costs of our Faraday synthesis algorithm is lower than for the 2+1D approach. Our method has led to the detection of a previously unidentified polarized source. Our bandwidth depolarization correction has been shown to accurately recover the intrinsic flux of depolarized sources, allowing us to image polarization data at a coarser frequency resolution while still maintaining high sensitivity at large Faraday depths. Future interferometers, such as MeerKAT+ and eventually the SKA, will require direction-dependent imaging to account for variations between antennas. In this context, direction-dependent Faraday synthesis is a promising approach for polarization studies. The code implementing this method will be made publicly available shortly after the publication of this paper as part of the \textsc{ddfacet} repository\footnote{\hyperlink{blue}{https://github.com/saopicc/DDFacet}}.

\begin{acknowledgements}

VG acknowledges support by the German Federal Ministry of Education and Research (BMBF) under grant D-MeerKAT III.
MB acknowledges funding by the Deutsche Forschungsgemeinschaft (DFG, German Research Foundation) under Germany's Excellence Strategy -- EXC 2121 ``Quantum Universe'' --  390833306 and the DFG research unit "Relativistic Jets"..
SPO acknowledges support from the Comunidad de Madrid Atracción de Talento program via grant 2022-T1/TIC-23797, and grant PID2023-146372OB-I00 funded by MICIU/AEI/10.13039/501100011033 and by ERDF, EU.
FdG acknowledges support from the ERC Consolidator Grant ULU 101086378.
We acknowledge data storage and computational facilities by the University of Bielefeld which are hosted by the Forschungszentrum Jülich and that were funded by German Federal Ministry of Education and Research (BMBF) projects D-LOFAR IV (05A17PBA) and D-MeerKAT-II (05A20PBA), as well as technical and operational support by BMBF projects D-LOFAR 2.0 (05A20PB1) and D-LOFAR-ERIC (05A23PB1), and German Research Foundation (DFG) project PUNCH4NFDI (460248186).

\end{acknowledgements}

%
%

\bibliography{ref}
\bibliographystyle{aa}

\begin{appendix}

\section{Bandwidth depolarization}
\label{sec:Bandwidth}

Bandwidth depolarization arises from the finite bandwidth of frequency channels, leading to signal averaging within each bin. This effect is visible as a function of $|\phi|$, where rapid Faraday rotation within a channel causes partial cancellation of Stokes $Q$ and $U$, reducing the observed polarization.

With our 3D deconvolution method, correcting for bandwidth depolarization is a requirement rather than an optional improvement, as neglecting it would leave significant residuals in the major cycle, preventing proper convergence. This is due to the different frequency resolutions in image space and data space, where less depolarization is experienced in the higher resolution data space. The approach is analogous to the time-dependent beam correction in \textsc{ddfacet}: while the residual image represents the time-averaged sky brightness distribution, the forward step applies time-dependent beam Jones matrices to ensure consistency with the visibility data. However, our correction does not require a facet-based approach, as the number of data points in $\phi$ is significantly smaller than the number of pixels in image space. Instead, we compute the correction by simulating channel averaging from degridding to gridding frequencies on synthetic samples with an RM corresponding to each sampled point in $\phi$. By performing RM synthesis on each sample and comparing the output flux to the input, we obtain a correction factor as a function of $|\phi|$.

\begin{figure}[H]
    \centering
    \includegraphics[width=\columnwidth]{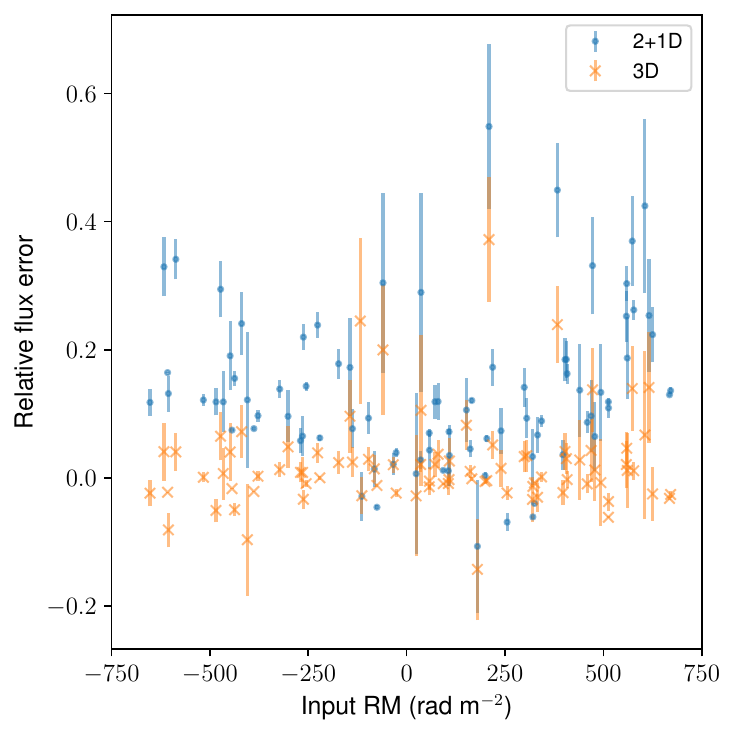}
    \caption{A comparison between the bandwidth depolarization corrected 2+1D (blue) and 3D (orange) methods on the effects of bandwidth depolarization. Only signals with a signal to noise above $8\sigma_{QU}$ are included in the figure.}
    \label{fig:Depolarization_corr}
\end{figure}

In an attempt to correct the 2+1D reconstruction for bandwidth depolarization, we used the \textit{bwdepol} module within \textsc{rm-tools}. The depolarization estimate is produced using the adjoint transform, discussed in \cite{Fine_2023}. Before we fit a Gaussian to the peaks in the deconvolved Faraday cubes, we divide the Faraday dispersion by the sensitivity estimate. The results are seen in Fig.~\ref{fig:Depolarization_corr}. We see that the correction is an overestimate of the true depolarization, as the errors appear to increase as a function of $|\phi|$.

\end{appendix}

\end{document}